\documentclass{ws-rv961x669}
\usepackage{ws-rv-van}     % numbered citation/references (default)
\usepackage{ws-rv-thm}     % comment this line when `amsthm / theorem / ntheorem` package is used
\usepackage{subfigure}     % required only when side-by-side / subfigures are used
\makeindex
\usepackage{xspace}
\usepackage{hyperref}
\usepackage{bbm}

\DeclareMathOperator{\re}{Re}
\DeclareMathOperator{\im}{Im}

\newcommand{\rep}[1]{\ensuremath\boldsymbol{#1}}

\newcommand{\Z}[1]{\ensuremath{\mathbbm{Z}_{#1}}} % z_N ->\Z{N}

\newcommand{\SU}[1]{\ensuremath{\mathrm{SU}(#1)}}
\newcommand{\SL}[1]{\ensuremath{\mathrm{SL}(#1)}}
\newcommand{\GL}[1]{\ensuremath{\mathrm{GL}(#1)}}

\newcommand{\I}{\mathrm{i}}

\newcommand{\CP}{\ensuremath{\mathcal{CP}}\xspace}
\newcommand{\x}{\ensuremath{\times}}

\begin{document}

\chapter[The flavor puzzle: textures and symmetries]{The Flavor Puzzle: Textures and Symmetries\label{ch:Nilles}}

\author[H.P. Nilles and S. Ramos-S\'anchez]{Hans Peter Nilles$^a$ and Sa\'ul Ramos-S\'anchez$^b$ \footnote{Talk given by HPN
at the ``Symposium in honor of Harald Fritzsch and his lasting contributions to Theoretical Physics'', July 14, 2023, 
LMU Munich, Germany.
Contribution to the ``Memorial volume with World Scientific in honor of Harald Fritzsch''.}}

%\author{Hans Peter Nilles and Saul Ramos-Sanchez]{\footnote{Author footnote.} 
%and Sa\'ul Ramos-S\'anchez\footnote{Author footnote.}}
%\index[aindx]{Author, F.} % or \aindx{Author, F.}
%\index[aindx]{Author, S.} % or \aindx{Author, S.}

\address{$^a$ Bethe Center for Theoretical Physics and Physikalisches Institut\\ 
der Universit\" at Bonn, Nussallee 12, 53115 Bonn, Germany\\
$^b$ Instituto de F\'isica, Universidad Nacional Aut\'onoma de M\'exico,\\
POB 20-364, Cd.Mx. 01000, Mexico}

\begin{abstract}
We discuss aspects of a promising top-down origin of flavor symmetries 
in particle physics. Modular transformations originating from string theory dualities are
shown to play a crucial role. We introduce the notion of an ``eclectic'' flavor scheme that
unifies traditional flavor symmetries, modular symmetries and \CP-transformations. It 
exhibits the phenomenon  of ``Local Flavor Unification'' with enhanced flavor symmetries
at fixed points or lines in moduli space. Successful fits of masses and mixing angles of
quarks and leptons are found in the vicinity of these points and lines. 
\end{abstract}

\body
%\tableofcontents
%%%%%%%%%%%%%%%%%%%%%%%%%%%%%%%%%%%%%%%%%%%%%%%%%%%%%%%%%%%%%%%%%%%%%%%%%%%%%%%%%%%
\section{Introduction}\label{sec:intro}

During my career I had many encounters with Harald Fritzsch. I still remember the first one at the University of Wuppertal in the 1970s. As students in Bonn, we became aware of the fact that Richard Feynman was supposed to give a colloquium at the Physics Department in Wuppertal. We, of course, all knew Feynman who was famous because of the ``Feynman Lectures on Physics''. So we decided to take our Volkswagen Beetle and drive with five people to Wuppertal. It was Harald Fritzsch, then a young professor at Wuppertal who had succeeded in bringing Feynman to Germany. Harald introduced Feynman and, as usual, said many good things about the speaker. Feynman started his talk by saying many good things about Harald as well and it became clear to us students that Harald had already made outstanding contributions to theoretical physics. Among those was the work on Colour and QCD (with Gell-Mann and Leutwyler~\cite{Fritzsch:1973pi}), SO(10) grand unification (with Minkowski~\cite{Fritzsch:1974nn}) and mass matrices and mixing angles of quarks and leptons~\cite{Fritzsch:1977vd,Fritzsch:1979zq}
to name just the most important ones.
In later years I had more encounters with Harald, first at CERN and later when I was a member of the physics department of the Technical University of Munich and the Max-Planck-Institute for physics in Freimann. Most recently, I met him during my stay at the LMU from Fall 2021 till spring 2022 with frequent visits of the ``Deeba'' restaurant in Barer Strasse where he was a regular guest. There, I learned more about Feynman, Gell-Mann and CalTech in the ``old times''.

In this talk I shall concentrate on the so-called ``flavor puzzle'', where Harald has made many influential contributions.

%%%%%%%%%%%%%%%%%%%%%%%%%%%%%%%%%%%%%%%%%%%%%%%%%%%%%%%%%%%%%%%%%%%%%%%%%%%%%%%%%%%
\section{The Flavor Puzzle}

Most parameters of the standard model of particle physics concern the flavor sector. This includes
\begin{enumerate}
\item the quark sector with 6 masses, 3 mixing angles and one phase, as well as

\item the lepton sector, again with 6 masses, 3 angles and one phase, and potentially additional parameters for Majorana neutrinos (as they appear for example in the 16-dimensional spinor representation of the SO(10) group of Fritzsch and Minkowski).
\end{enumerate}
The pattern of those parameters is quite peculiar. In the quark sector we have hierarchical masses and small mixing angles. In the lepton sector we have two large and one small mixing angle, again a hierarchical mass pattern and extremely small neutrino masses. The flavor structure of quarks and leptons is thus very different.
The ``flavor puzzle'' aims at an understanding of these masses and mixing angles as well as possible relations among them. In the bottom-up approach the goal was a description with very few parameters. Historically, this would then lead to predictions for those masses and mixing angles that were not yet measured at that time.

One approach was to find so-called ``mass textures'', an ansatz for a mass matrix with very few parameters, some of them even zero (so-called ``texture zeros''). The simplest example~\cite{Fritzsch:1977za,Weinberg:1977hb} 
is the description of masses of down- and strange-quark masses as well as the Cabibbo angle: the ansatz 
\begin{equation}
\begin{pmatrix}
0 & m\\
m & M 
\end{pmatrix}
\rightarrow
\begin{pmatrix}
m^2/M & 0\\
0 & M 
\end{pmatrix}
=
\begin{pmatrix}
m_d & 0\\
0 & m_s
\end{pmatrix}
\end{equation}
with $m\ll M$. Two parameters $m, M$ describe two masses and one mixing angle (the Cabibbo angle),
\begin{equation}
 \sin\theta_C \sim \tan \theta_C = \sqrt{(m_d/m_s)} \  ,
\end{equation}
a quite successful description. Of course, with three families of quarks and leptons the situation becomes more complicated. Fritzsch had a successful suggestion with four ``texture zeros'' in the so-called ``Fritzsch mass matrix''~\cite{Fritzsch:1977vd}
\begin{equation}
\begin{pmatrix}
0 & a & 0\\
a^* & 0 & b\\
0 & b^* & c
\end{pmatrix}\,,
\end{equation}
which leads to various relations between masses and mixing angles. For a summary of this approach, we refer to the work of Fritzsch and Xing~\cite{Fritzsch:1999ee}. 
In this bottom-up approach one describes the various hierarchical structures of quark and lepton masses by fitting existing data. After having identified a successful texture, one would then try to understand its origin from an underlying principle. Such an understanding could come as a result of symmetries.  The idea of possible relations between textures and symmetries leads still in our days to interesting results~\cite{Kikuchi:2022svo}.

Throughout the years, discrete symmetries have been used extensively in the discussion of flavor structure. There have been many fits from a bottom-up perspective with various discrete symmetries ($S_{3}, A_{4}, S_{4}, A_{5}, \Delta(27), \Delta(54)$ and many more~\cite{Ishimori:2010au}). Flavor symmetries seem to require different models for quark and lepton sectors, hierarchical structure can appear in the case of slightly broken symmetries. This bottom-up model building leads to many reasonable fits for various choices of groups and representations. Maybe too many successful fits. We need to have a top-down explanation of flavor to clarify the flavor puzzle.

%%%%%%%%%%%%%%%%%%%%%%%%%%%%%%%%%%%%%%%%%%%%%%%%%%%%%%%%%%%%%%%%%%%%%%%%%%%%%%%%%%%%%%%
\section{Discrete Flavor Symmetries}

In our top-down approach we consider string theory with compactified extra dimensions. Discrete symmetries can appear as a result of the geometry of extra dimensions and the geography of fields localised in compact space. We encounter two types of symmetry: traditional and modular flavor symmetry. The traditional flavor symmetry can already be found in quantum field theory of point particles. These symmetries are linearly realised and need to be broken spontaneously by vacuum expectation values of so-called flavon fields. Another type of discrete flavor symmetries are modular symmetries that originate from string theory dualities. Originally they have been discovered within the framework of orbifold compactifications~\cite{Lauer:1989ax,Lauer:1990tm} 
via conformal world sheet calculations. The application to flavor physics in the lepton sector has been pioneered by Feruglio~\cite{Feruglio:2017spp}.
They are non-linearly realised in the low-energy field theory description. In separate schemes, traditional and modular flavor symmetries are known to lead to interesting results~\cite{Feruglio:2019ybq}. In our top-down framework, they appear combined, forming
the so-called ``Eclectic Flavor Group''~\cite{Nilles:2020nnc,Nilles:2020kgo,Nilles:2020tdp,Nilles:2020gvu}. 
Inspired by this, bottom-up eclectic~\cite{Ding:2023ynd} and quasi-eclectic~\cite{Chen:2021prl} models with promising phenomenology have been constructed.)

Thus, the discrete flavor symmetries arise as a consequence of the ``string geometry'' of extra dimensions. Strings are extended objects and this reflects itself in generalised aspects of geometry that include the winding modes of strings. They share the traditional symmetries of quantum field theories of point particles and have additional modular or symplectic flavor symmetries as a result of string duality transformations that exchange winding and momentum modes. In the following we illustrate this in a simple example with localised matter fields on a twisted two-dimensional torus, relevant for string theory compactifications with elliptic fibrations.

\begin{figure}
\centerline{\includegraphics[width=10.5cm]{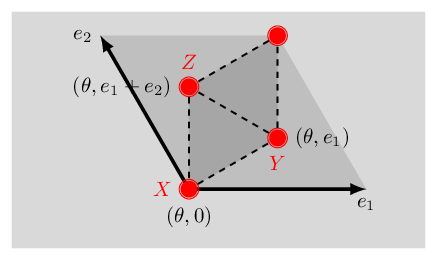}}
\caption{The $\mathbb{T}^2/\Z3$ orbifold. The points of the upper half plane are identified 
when they differ by lattice vectors $e_1$ and $e_2$ (which define a torus (light shaded area)) 
or a rotation by 120 degrees (which define the orbifold (dark shaded area)). There are three fixed points,
$X, Y$ and $Z$. The geometrical object is obtained by folding the dark shaded area along the line between
$Y$ and $Z$, fliping the triangle over and gluing at the edges. }
\label{fig:Z3}
\end{figure}

Let us start with the traditional flavor symmetries of the $\mathbb{T}^2/\Z3$ orbifold shown in fig.~\ref{fig:Z3}.
In string theory discrete symmetries can arise from geometry and string selection rules. Consider twisted fields $X, Y, Z$ on the fixed points of the orbifold. There is an 
$S_3$-symmetry from the interchange of fixed points and a $\Z{3}\x\Z{3}$ symmetry from string theory selection rules. The full discrete flavor group is $\Delta(54)$, resulting from the multiplicative closure of $S_3$ and $\Z{3}\x\Z{3}$~\cite{Kobayashi:2006wq}.
This symmetry has 54 elements and is a non-abelian subgroup of \SU3. The twisted states $(X, Y, Z)$ transform as triplets under $\Delta(54)$. This group could be interpreted as the traditional flavor symmetry of three families of quarks (or leptons). 

To understand modular flavor symmetry, let us first consider a warm-up example in one dimension: a circle of radius $R$ (or a particle in a box with periodic boundary conditions). There is a discrete spectrum of momentum modes (Kaluza-Klein modes) whose density is governed by $m/R$ (with $m$ integer). Heavy modes decouple for $R \to 0$. Now consider a string. We have Kaluza-Klein modes $m/R$ as before. In addition, strings can wind around the circle and the spectrum of winding modes is governed by $nR$ ($n$ integer). A tower of light modes appears in the limit $R \to 0$. This interplay of winding and momentum modes is the origin of so-called T-duality, where one simultaneously interchanges winding and momentum modes as well as $R$ and $1/R$. This transformation maps a theory to its T-dual theory. At the self-dual point $R^{2} = 1$, we have a \Z2 symmetry.

\begin{figure}
\centerline{\includegraphics[width=8.5cm]{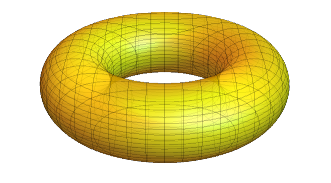}}
\caption{The two-torus $\mathbb{T}^2$ exhibiting two cycles for winding modes.}
\label{fig:donut}
\end{figure}

\begin{figure}
\centerline{\includegraphics[width=12.5cm]{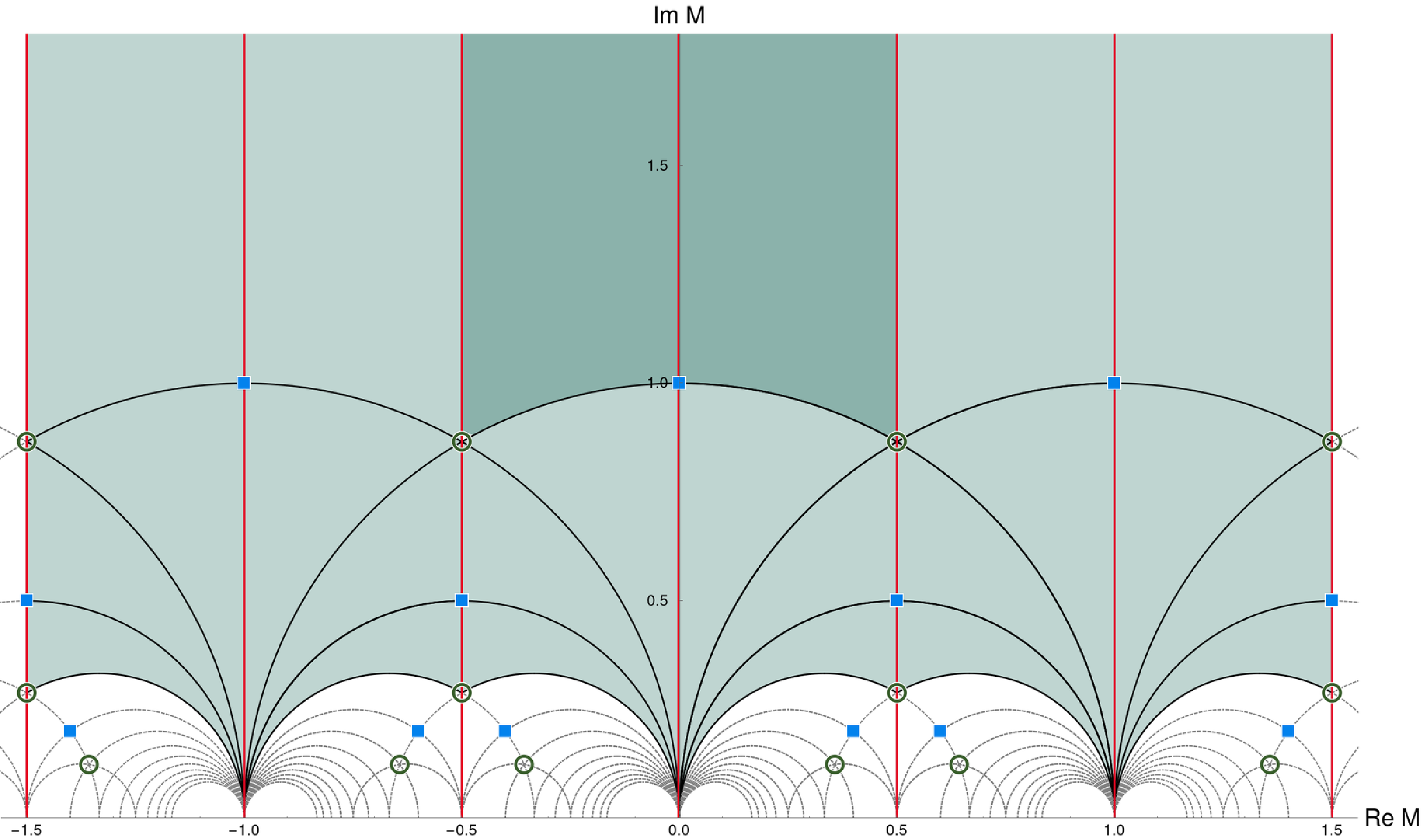}}
\caption{ The fundamental domain of the modular group \SL{2,\Z{}} is the dark shaded area (extending to $\I\infty$).
There are three fixed points at $M=\I, \omega=\exp(2\pi\I/3)$ and $\I\infty$.
The light shaded area is the fundamental domain of $\Gamma(3)=\SL{2,3\Z{}}$.}
\label{fig:fundamentalDomain}
\end{figure}

In case of two dimensions, we have a torus and the string can wind around two different cycles (fig.~\ref{fig:donut}).
This torus can be described by a complex modulus $M$ with values in the complex upper half plane. The generalisation of the above T-duality to two dimensions leads to modular transformations that exchange winding and momentum-modes and act nontrivially on the moduli of the torus. These transformations build
the modular group $\SL{2,\Z{}}$, which is generated by two elements
\begin{equation}
{\rm S},\ {\rm T}:\ \ \ \ {\rm with} \ \  {\rm S}^4 = ({\rm ST})^3 = 1 \ \ \text{and}\ \  {\rm S}^2 {\rm T} = {\rm T}\, {\rm S}^2\,.
\end{equation}
A modulus $M$ transforms as
\begin{equation}
{\rm S}:\ \ M\rightarrow -{1\over M}\qquad {\rm and}\qquad{\rm  T}: M\rightarrow M+1\,.
\end{equation}
Further symmetry transformations might include 
\begin{equation}
{\rm U}:\ \ M\to -\overline{M}
\end{equation}
that can be the source of a \CP  or \CP-like symmetry. The fundamental domain of the modular group is given in fig.~\ref{fig:fundamentalDomain}.
All points in the upper half plane can be mapped to the fundamental domain via a \SL{2,\Z{}}  transformation. An element $\gamma$ of \SL{2,\Z{}} acts on the modulus as
\begin{equation}
\gamma: \ \ M\rightarrow \frac{aM+b }{cM+d} 
\end{equation}
with $ad-bc=1$ and integer $a, b, c, d$. Matter fields transform as modular functions of weight $k$
\begin{equation}
\gamma: \ \ \ {\phi
 \rightarrow (cM+d)^{k} \rho(\gamma)\phi 
 \ ,
}
\end{equation}
where $\rho(\gamma)$ is a unitary representation of $\gamma$, belonging to a finite modular group $\Gamma_N$ or $\Gamma_N'$.
Yukawa couplings transform as modular functions too. In a supersymmetric theory the superpotential ($W$) might transform nontrivially as well~\cite{Lauer:1989ax,Lerche:1989cs}.
The combination $G=K+\log(|W|^2)$ (with K\"ahler potential $K$) must be invariant under \SL{2,\Z{}}~\cite{Ferrara:1989qb}.

Let us now look at the modular symmetries of the $\mathbb{T}^2/\Z3$ twisted torus. To allow for the twist, the lattice vectors $e_1$ and $e_2$ of fig.~\ref{fig:Z3} must have the same length and the angle is fixed to 120 degrees. The relevant modular transformations form a subgroup of \SL{2,\Z{}}: $\Gamma(3)=\SL{2,3\Z{}}$, where the shift by an integer $n$ in \SL{2,\Z{}} is replaced by the shift $3n$.
 $\Gamma(3)$ is a mod(3) subgroup of \SL{2,\Z{}}, still a group with infinitely many elements. Taking the quotient, 
we obtain the discrete modular flavor group $\Gamma_3'=\SL{2,\Z{}}/\Gamma(3)$. This is a group with 24 elements known as $T'\cong\SL{2,3}$. It is the double cover of 
$\Gamma_3\cong A_4$  (the group of even permutations of 4 elements). $\Gamma_3\cong A_4$ acts on the modulus. $\Gamma_3'\cong T'$ is the double cover which also acts nontrivially on the twisted fields. (This is a similar 
situation as in the Lorentz group where bosons transform under $\mathrm{O}(3)$ while the double cover \SU2 is needed for fermions). The \CP-transformation ${\rm U}: M\to -\overline M$ completes the picture and enhances the discrete modular group to \GL{2,3}, a group with 48 elements.
This provides an example that illustrates the two types of flavor groups
\begin{enumerate}
\item traditional flavor group: $\Delta(54)$, and

\item modular flavor group: $T'$.
\end{enumerate}
The eclectic flavor group is defined as the multiplicative closure of these groups. Here we obtain $\Omega(1) \cong [648, 533]$ from $\Delta(54)$ and $T'$ as well as [1296, 2891] if we include \CP~\cite{Baur:2019kwi,Novichkov:2019sqv}. (We use the notation of the SmallGroup (SG) library of GAP~\cite{GAP4}, where the first number in the squared parentheses is the order of the group.)
The eclectic group is the largest possible flavor group of a given system, but it is not necessarily linearly realised.

%%%%%%%%%%%%%%%%%%%%%%%%%%%%%%%%%%%%%%%%%%%%%%%%%%%%%%%%%%%%%%%%%%%%%%%%

\section{Local Flavor Unification}

\begin{figure}[t!]
\centerline{\includegraphics[width=12.5cm]{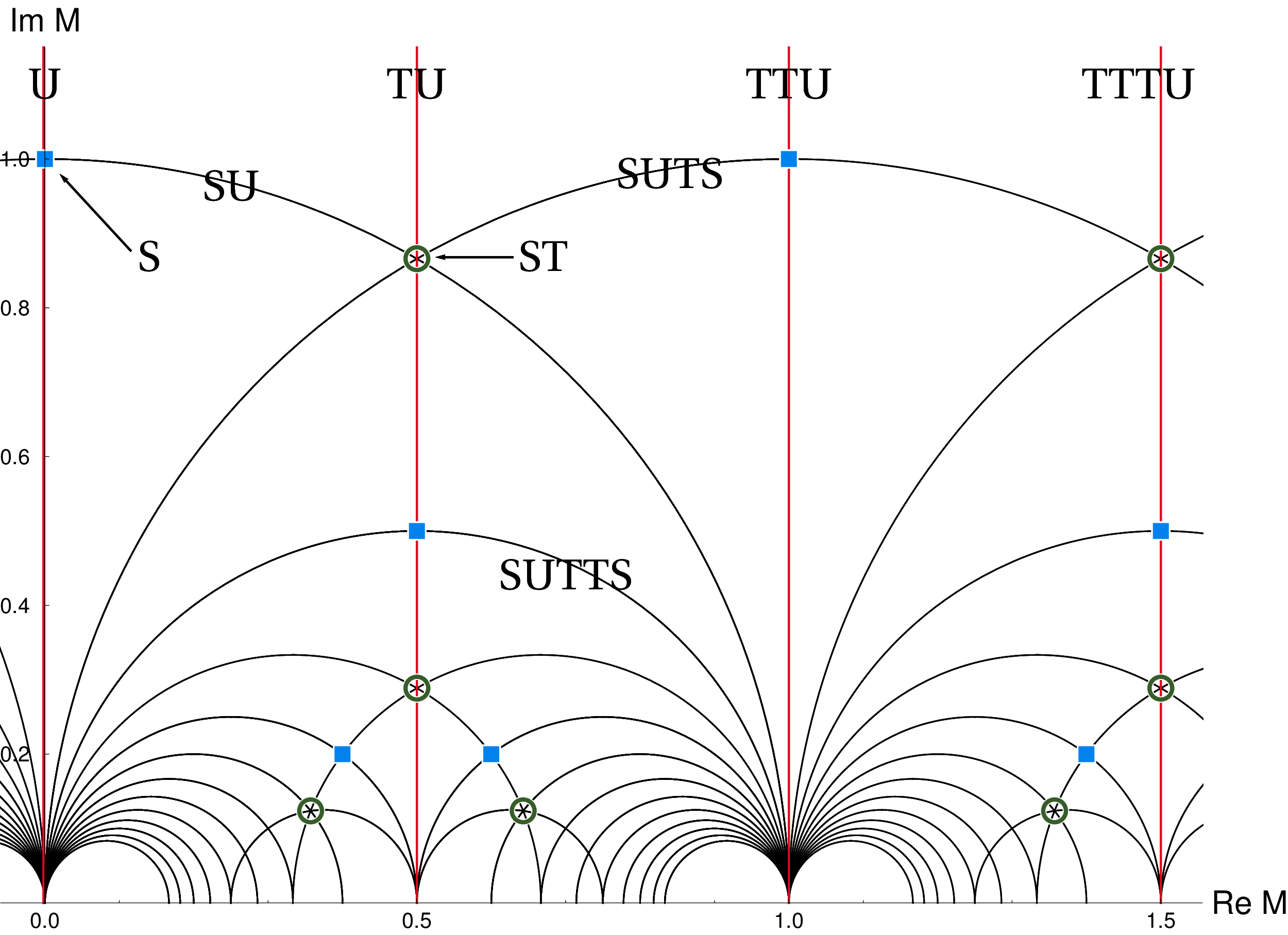}}
\caption{Fixed points and lines under the transformations ${\rm S}: M\to -1/M$, ${\rm T}: M\to M+1$ and ${\rm U}: M\to -\overline M$.}
\label{fig:fixedLoci}
\end{figure}

\begin{figure}
\centerline{\includegraphics[width=12.5cm]{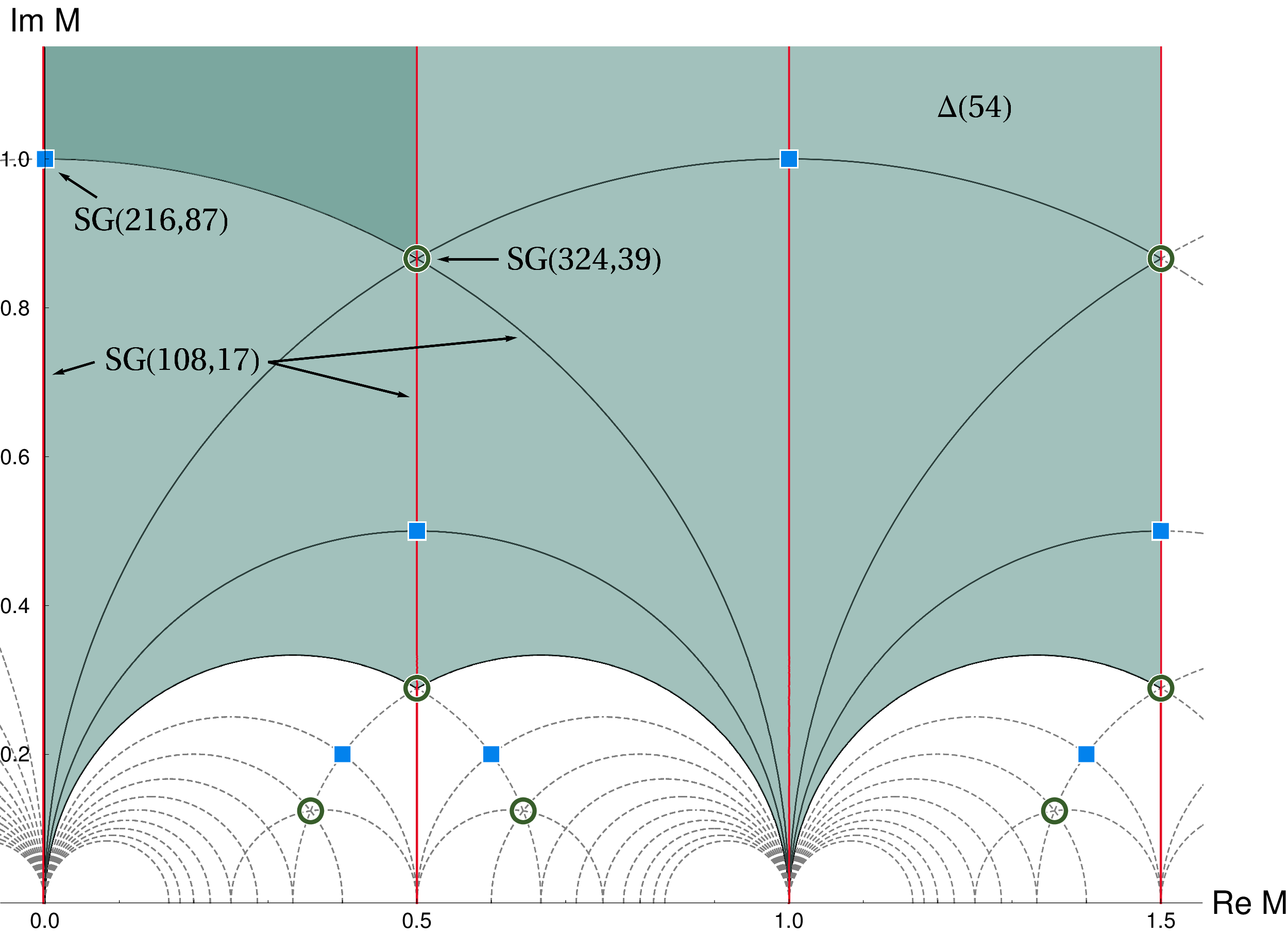}}
\caption{The fundamental domain of \GL{2,3\Z{}} with fixed points and fixed lines exhibiting the phenomenon of
``Local Flavor Unification''. The traditional flavor group $\Delta(54)$ is universal in moduli space. Symmetry 
enhancements to groups with 108 elements are along fixed lines. When two fixed lines meet, the enhancement
gives groups with 216 elements and 324 elements when three lines meet.}
\label{fig:Z3symmetries}
\end{figure}

Let us look at the moduli space of $\Gamma(3)=\SL{2,3\Z{}}$ (as shown in fig.~\ref{fig:fundamentalDomain}, light shaded area). It is larger than
that of \SL{2,\Z{}}.
If we include the ${\rm U}$ transformation $M\to -\overline M$, this moduli space is reduced to the one with $\re(M)\geq 0$. Under the transformations ${\rm S}, {\rm T}$ and ${\rm U}$, we find a variety of fixed lines and points as shown in fig.~\ref{fig:fixedLoci}.
At this fixed loci we have enhanced symmetries (as in the case of $R^{2} = 1$ of our one-dimensional example). These symmetries are shown in fig.~\ref{fig:Z3symmetries}.
At a generic point in moduli space we have the traditional flavor group $\Delta(54)$, i.e.\ it is universal in moduli space (while the modular flavor group is non-linearly realised). On the fixed lines the symmetry is enhanced to a group with 108 elements [108,17]. When two lines meet we have the group [216,87], and when three lines meet the group [324,39]. The largest linearly realised group in this example has thus 324 elements, while the eclectic group is [1296,2891].

The modular flavor group is non-universal in moduli space.
This leads to a phenomenon we call ``Local Flavor Unification''~\cite{Baur:2019kwi,Baur:2019iai}.
At special locations in moduli space we have an enhanced flavor symmetry. Some of the flavor parameters, such as masses or mixing angles, might vanish at these points and lines. At the red vertical lines of fig.~\ref{fig:Z3symmetries} we have e.g.\ unbroken \CP. If we are close to the fixed points or lines, some of the parameters are thus small and we can explain mass hierarchies or small mixing angles. 

This phenomenon of ``Local Flavor Unification'' is a prediction of string theory. It provides traditional flavor symmetries which are universal in moduli space as well as modular flavor symmetries and \CP that are non-universal and lead to local flavor enhancements. They unify in the eclectic picture of flavor symmetry. Both have to be present, you cannot just have one without the other. The non-universality in moduli space leads to local flavor unification at specific points in moduli space with hierarchical structures of masses, mixing angles and phases in the vicinity of fixed points or lines.

%%%%%%%%%%%%%%%%%%%%%%%%%%%%%%%%%%%%%%%%%%%%%%%%%%%%%%%%%%%%%%%%%%%%%%%%%%%%%%%%%%
\section{Top-Down Model Building}

\begin{figure}[t!]
\centerline{\includegraphics[width=12.5cm]{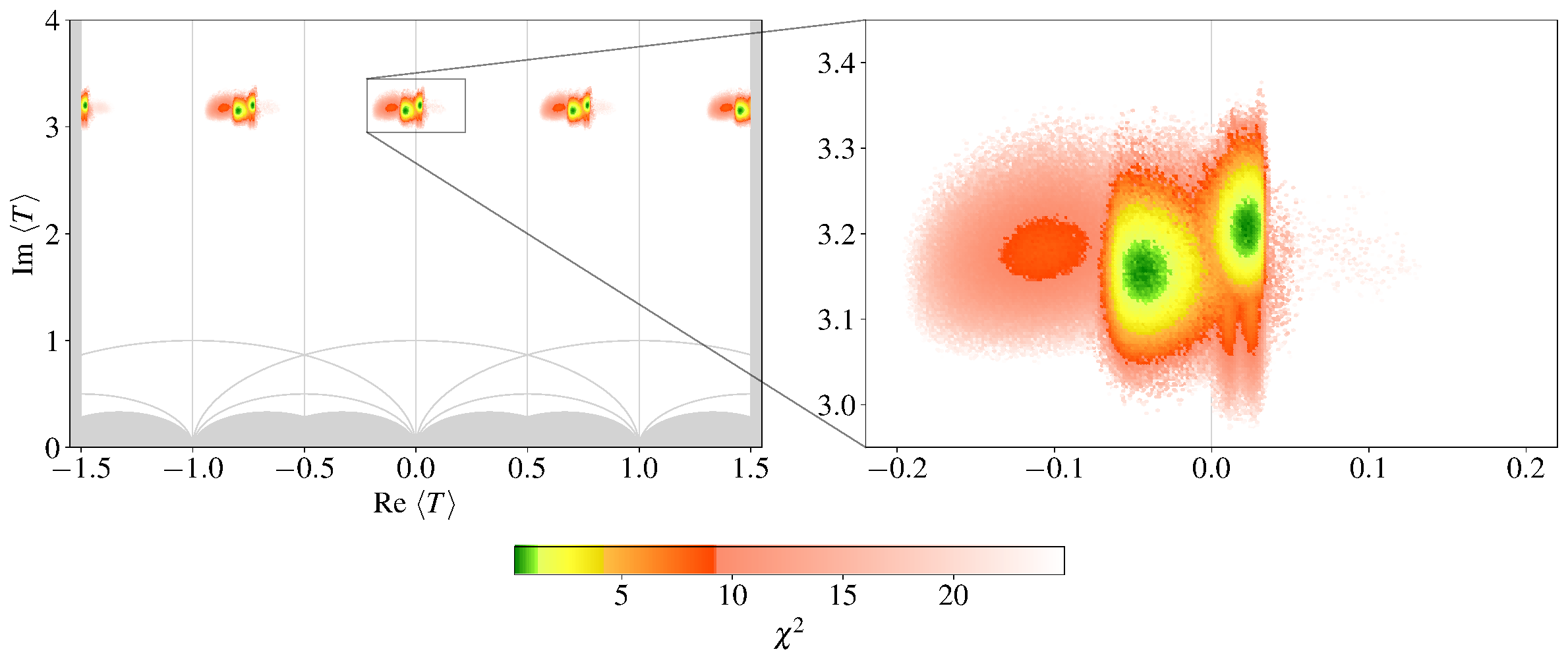}}
\caption{The best-fit modulus is at $\re(T)$ close to zero and at $\im(T)\sim 3$ (which for all practical purposes 
is close to $\infty$).}
\label{fig:bestfit-modulus}
\end{figure}

\begin{figure}
\centerline{\includegraphics[width=12cm]{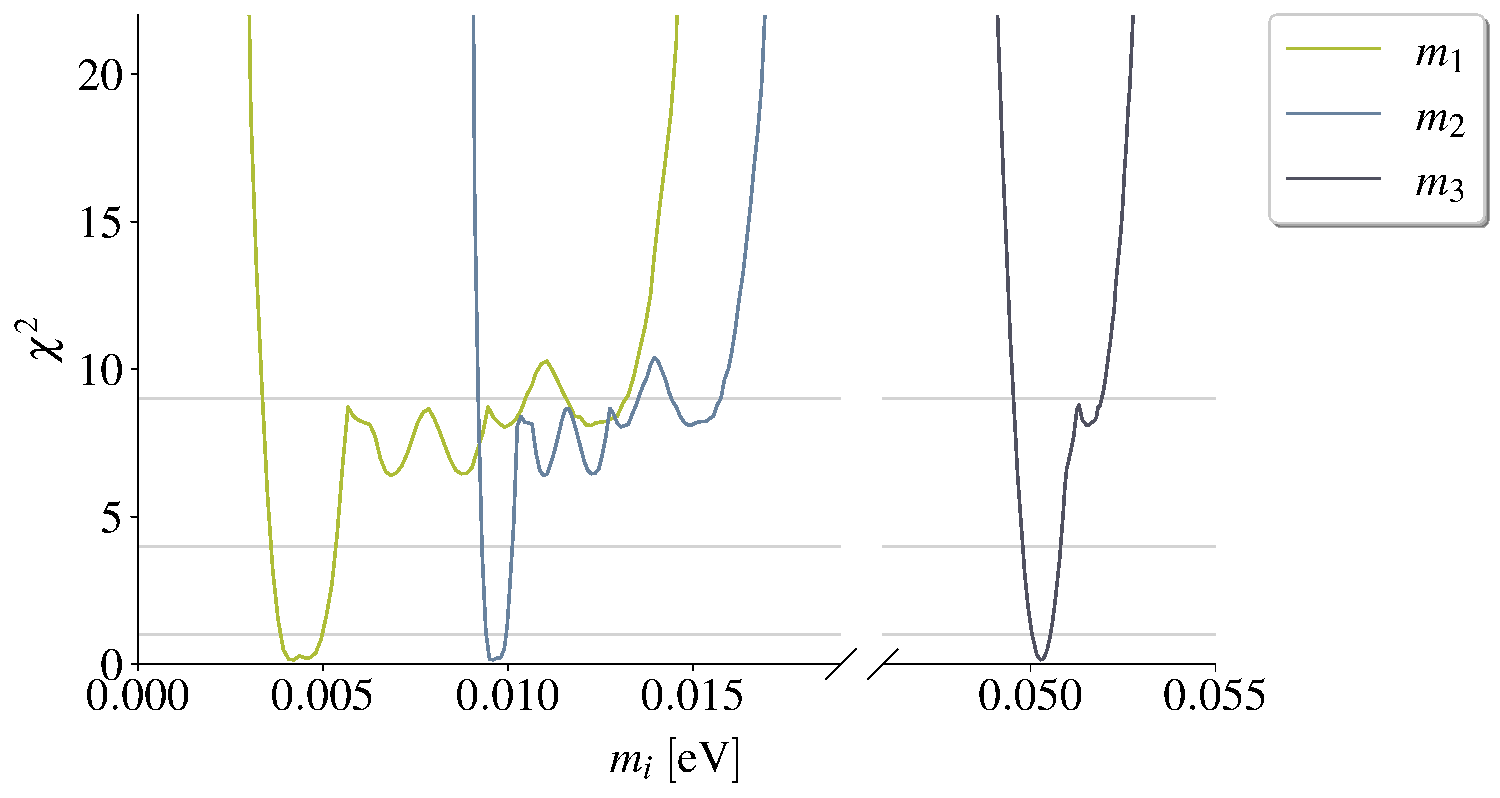}}
\caption{The model predicts a normal hierarchy for the neutrino masses $m_i$.}
\label{fig:neutrino}
\end{figure}

\begin{figure}
\centerline{\includegraphics[width=12cm]{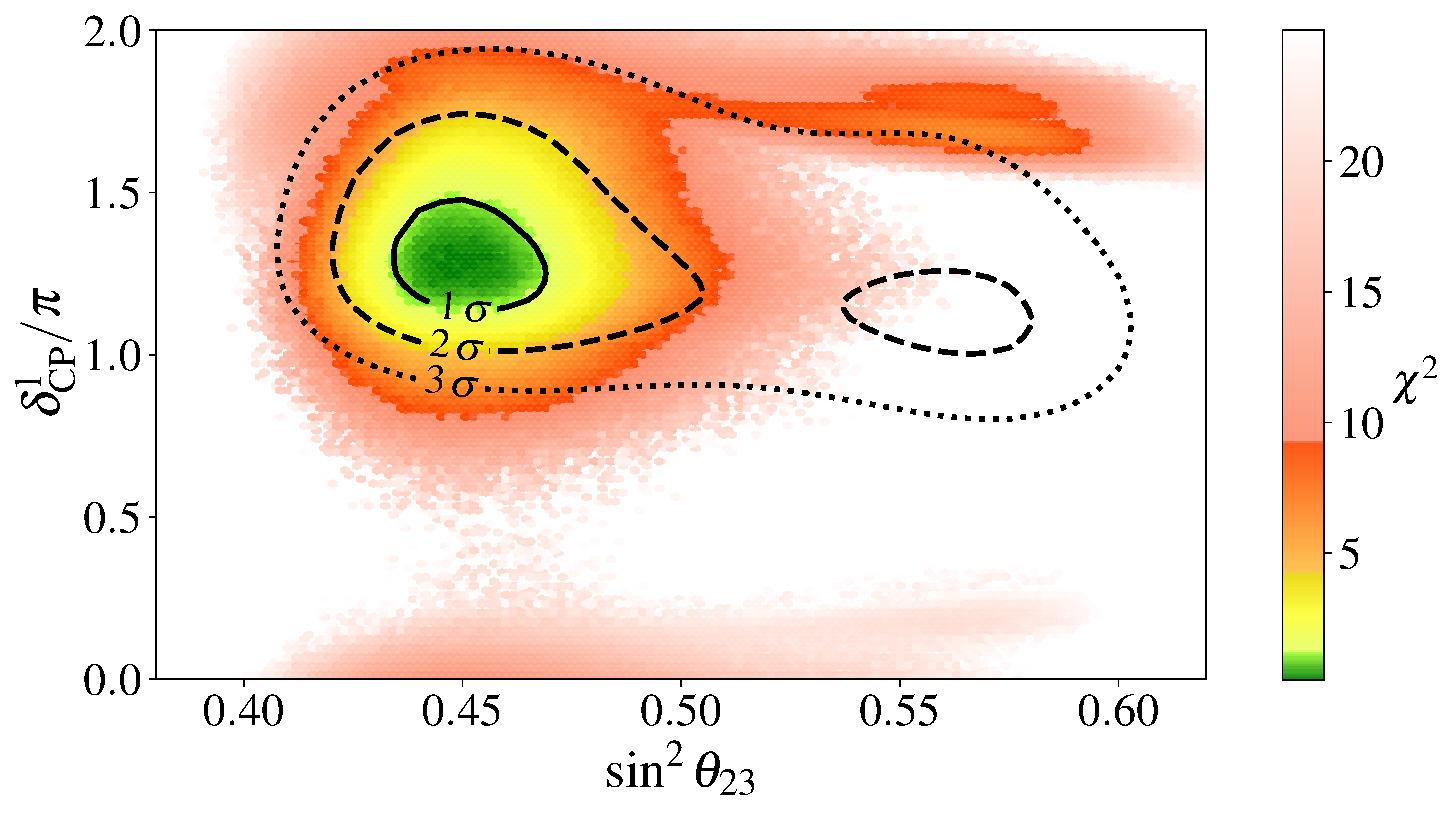}}
\caption{Best-fit values for the mixing angle $\theta_{23}$ versus \CP-angle.}
\label{fig:mixing-CP}
\end{figure}

\begin{figure}
\centerline{\includegraphics[width=12.5cm]{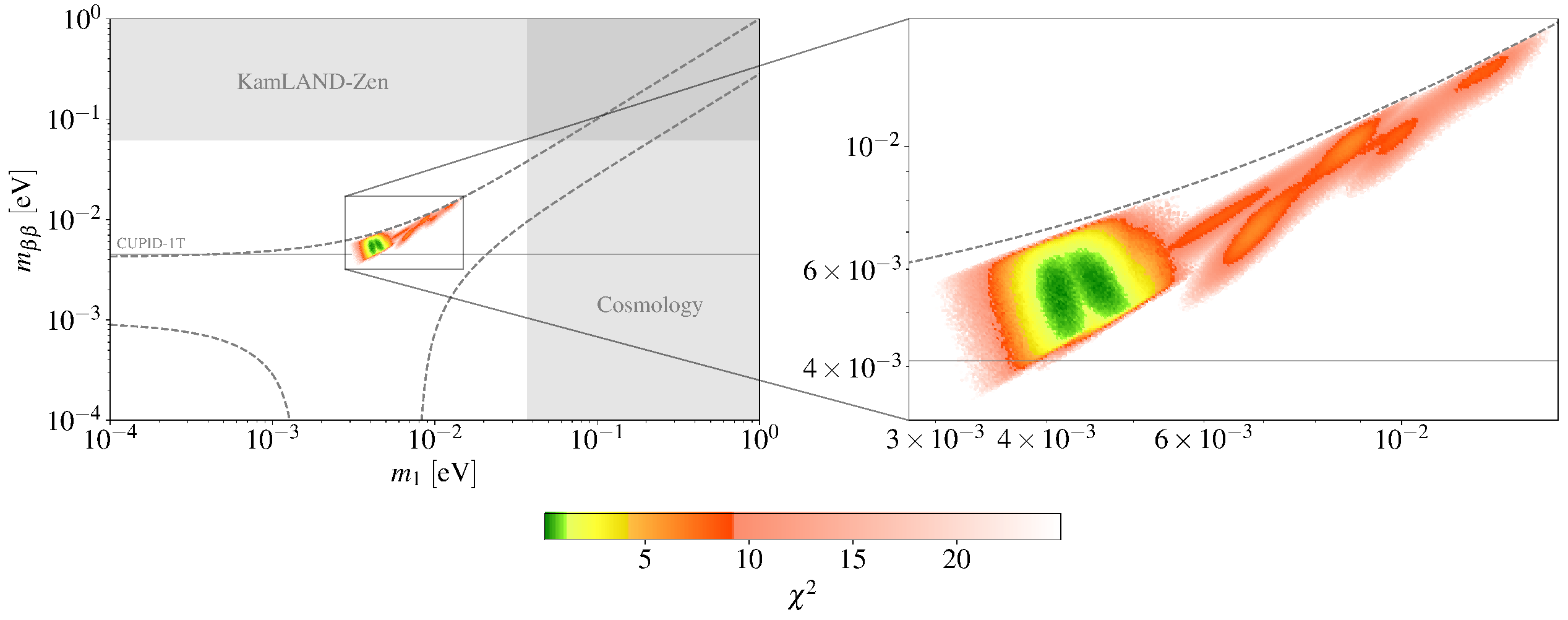}}
\caption{The model predicts a rather high value for the effective Majorana mass of the neutrinos, here in a plot
versus the mass of the lightest neutrino $m_1$.}
\label{fig:Majorana}
\end{figure}

So far, we have discussed the symmetries of a 2-dimensional compact space. In realistic string theory we have a 6-dimensional compact space and the two-torus has to be suitably embedded therein. In general, this will enhance the eclectic flavor group by additional $R$-symmetries~\cite{Kobayashi:2004ya,Nilles:2013lda,CaboBizet:2013gns,Nilles:2017heg,Nilles:2020gvu}. 
As a starting point, we shall consider orbifold compactifications of the heterotic string on a $\mathbb{T}^6/(\Z3\x\Z3)$-orbifold, as discussed by Carballo-P\'erez, Peinado and Ramos-S\'anchez~\cite{Carballo-Perez:2016ooy}.
They constructed realistic models with standard model gauge group and three families of quarks and leptons and traditional flavor group $\Delta(54)$. We selected a specific model~\cite{Baur:2022hma} where the twisted states are in a $\rep3_2$-representation of $\Delta(54)$ as well as a $\rep1\oplus\rep2'$ of $T'$ with modular weight $k=-2/3$. Observe, that in the top-down approach these representations and the modular weight are restricted to specific values. This is in contrast to bottom-up constructions where the representation of the modular group and the modular weights can be freely chosen. Our model contains flavon fields needed for a breakdown of the traditional flavor group $\Delta(54)$. The interplay of this breakdown and the breakdown of $T'$ via a modulus in the vicinity of a fixed point allows a hierarchy of masses and mixing angles as expected from the phenomenon of ``Local Flavor Unification''~\cite{Baur:2019kwi,Baur:2019iai}. 
Uncontrollable corrections to the K\"ahler-potential~\cite{Chen:2019ewa}
are avoided through the approximate traditional flavor symmetry $\Delta(54)$, while controllable corrections allow for a description of the quark sector~\cite{Baur:2022hma}. 
The model predicts a see-saw mechanism in the lepton sector and a normal hierarchy for neutrino masses. The best-fit modulus, shown in fig.~\ref{fig:bestfit-modulus} is at $\re(M)$ close to zero and $\im(M)\sim 3$ (which for all practical purposes is close to $M=\I\infty$).
The neutrino mass spectrum is given in fig.~\ref{fig:neutrino}
and mixing angles versus \CP-phase given in fig.~\ref{fig:mixing-CP}. 
The model predicts a rather high value for the effective Majorana masses (fig.~\ref{fig:Majorana}). 
So far, models with traditional flavor symmetry $\Delta(54)$ and discrete modular group $T'$ seem to be good candidates to address the flavor problem. There are numerous bottom-up constructions with these groups as well as realistic string models from \Z3-orbifolds. We need, however, more top-down constructions to clarify the situation and make contact with specific bottom-up constructions. Please find more information about bottom-up constructions in a recent review.\cite{Kobayashi:2023zzc}.

%%%%%%%%%%%%%%%%%%%%%%%%%%%%%%%%%%%%%%%%%%%%%%%%%%%%%%%%%%%%%%%%%%%%%%%%%%%%%%%%%%%%
\section{Messages and Open Questions}

\begin{figure}
\vskip-0.4cm
\centerline{\includegraphics[width=12.5cm]{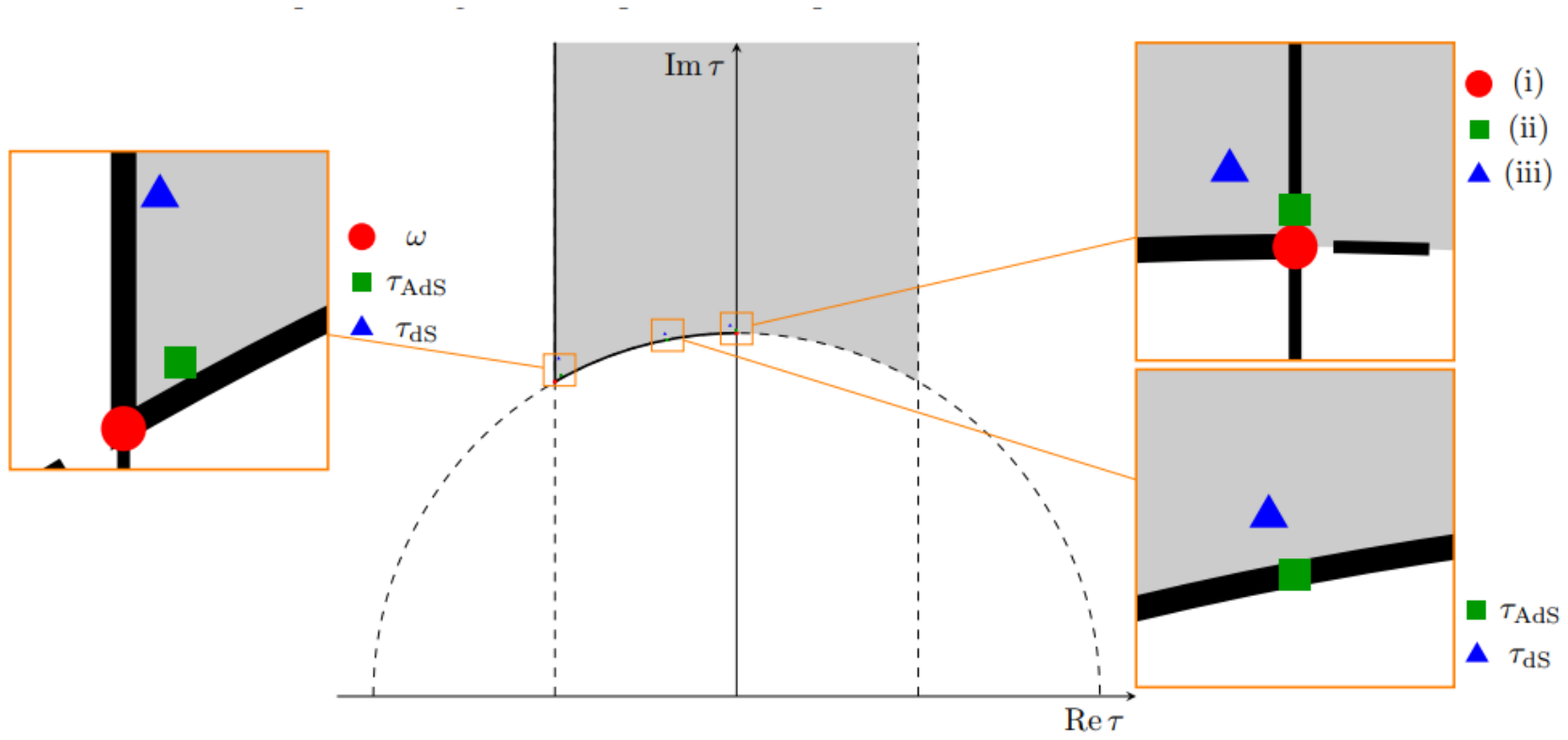}}
\vskip-1.8cm
\caption{Location of minima of the scalar potential for the modulus $\tau$. The red circles denote the fixed points, green squares AdS-vacua and blue triangles dS minima.}
\label{fig:minima}
\end{figure}

The top-down approach to flavor symmetries leads to a unification of traditional discrete flavor, \CP and modular symmetries within an electric flavor scheme. The modular symmetry is a prediction of string theory that is always accompanied by a traditional flavor symmetry. \CP is a consequence of the symmetries of the underlying string theory. There are non-universal symmetry enhancements in moduli space that lead to the phenomenon of ``Local Flavor Unification''. Spontaneous symmetry breakdown can be understood as a motion in moduli space. 

From the top-down approach, up to now, phenomenology of eclectic flavor symmetries has only been studied in models based on the \Z3 orbifold.
It remains to be seen whether similar results can be obtained for other cases like \Z{k} with $k=2,4,6$ or generalisations based on the Siegel modular group~\cite{Ding:2020zxw,Baur:2020yjl}. In addition, it would be interesting to explore the possibility of this eclectic scheme in other top-down scenarios~\cite{Kobayashi:2018rad,Ohki:2020bpo,Kikuchi:2020frp}, where metaplectic flavor symmetries are known to appear~\cite{Almumin:2021fbk} as modular component.

In agreement with the expectations from ``Local Flavor Unification'', it has been observed that most of the successful fits in the bottom-up approach predict the modulus in the vicinity of fixed points and fixed lines~\cite{Feruglio:2021dte,Wang:2021mkw,Kikuchi:2022svo,Feruglio:2022koo,Petcov:2022fjf,Petcov:2023vws}.
The question remains whether there exists a dynamical mechanism that drives the modulus towards these regions. This is a question of moduli stabilisation and the identification of the minima of the scalar potential of the theory. It has long been known that the modular symmetry favours extrema at the boundary of moduli space~\cite{Cvetic:1991qm} 
(with some exceptions for minima close to $M=\omega$~\cite{Dent:2001ut,Novichkov:2022wvg,Leedom:2022zdm}). 
At these boundaries of moduli space, however, we have enhanced symmetries and this does not allow for a satisfactory fit of masses and mixing angles. Fortunately, a closer inspection of the potential reveals the fact that all these minima have negative vacuum energy (AdS). We then need an ``up-lift'' to obtain vacua with vanishing (or slightly positive) vacuum energy. Such an up-lift via ``matter superpotentials''~\cite{Lebedev:2006qq}~has been discussed recently~\cite{Knapp-Perez:2023nty} and it has been shown that it moves the minima slightly away from the boundary as shown in fig.~\ref{fig:minima}. 
This provides a dynamical mechanism to obtain realistic minima in the vicinity of fixed points and lines as desired for realistic fits of masses and mixing angles of quark and leptons.

We have seen that string theory provides us with all the necessary ingredients to attack the flavor problem: traditional flavor groups, discrete modular groups and a natural candidate for \CP. The non-universality of flavor symmetry in moduli space leads to the phenomenon of ``Local Flavor Unification'' including a dynamical mechanism that drives the modulus to the vicinity of fixed points and lines of moduli space. This opens up a new arena for flavor model building.

\section*{Acknowledgments}
It is a pleasure to thank Alexander Baur, V\'ictor Knapp-P\'erez, Xiang-Gan Liu, Michael Ratz, Andreas Trautner, and Patrick K.S.\ Vaudrevange for fruitful and enjoyable collaborations.
The work of SR-S was partly supported by UNAM-PAPIIT IN113223, CONACyT grant CB-2017-2018/A1-S-13051, and Marcos Moshinsky Foundation.

%\bibliographystyle{ws-rv-van}
%\bibliography{Nilles}

%\printindex[aindx]                 % to print author index
%\printindex                         % to print subject index
\end{document}